\documentclass[%
 reprint,
 amsmath,amssymb,
 aps,
superscriptaddress,
]{revtex4-1}

\usepackage{graphicx}
\usepackage{dcolumn}
\usepackage{bm}
\usepackage{multirow}
\usepackage{array}
\usepackage{bm}
\usepackage{color}

\begin{document}

\title{Magnetism and   {weak} electronic correlations in Kagome metal ScV$_6$Sn$_6$}

\author{Tianye Yu}
\affiliation{Shenyang National Laboratory for Materials Science, Institute of Metal Research,Chinese Academy of Sciences, Shenyang 110016, China.}

\author{Junwen Lai}
\affiliation{Shenyang National Laboratory for Materials Science, Institute of Metal Research,Chinese Academy of Sciences, Shenyang 110016, China.}
\affiliation{School of Materials Science and Engineering, University of Science and Technology of China, Shenyang 110016, China.}

\author{Xiangyang Liu}
\affiliation{Shenyang National Laboratory for Materials Science, Institute of Metal Research,Chinese Academy of Sciences, Shenyang 110016, China.}
\affiliation{School of Materials Science and Engineering, University of Science and Technology of China, Shenyang 110016, China.}

\author{Peitao Liu}
\affiliation{Shenyang National Laboratory for Materials Science, Institute of Metal Research,Chinese Academy of Sciences, Shenyang 110016, China.}

\author{Xing-Qiu Chen}
\email{xingqiu.chen@imr.ac.cn}
\affiliation{Shenyang National Laboratory for Materials Science, Institute of Metal Research,Chinese Academy of Sciences, Shenyang 110016, China.}

\author{Yan Sun}
\email{sunyan@imr.ac.cn}
\affiliation{Shenyang National Laboratory for Materials Science, Institute of Metal Research,Chinese Academy of Sciences, Shenyang 110016, China.}



\begin{abstract}
As one class of typical quantum materials,
Kagome metals in $A$V$_3$Sb$_5$($A$ = K, Rb, Cs) have attracted extensive 
attentions due to their interesting physical properties and different 
quantum phases of charge density wave (CDW), superconductivity and 
nontrivial topology. Recently, a new CDW phase in ScV$_6$Sn$_6$ was 
experimentally observed and inspired a wide study of the mechanism of
driving force. 
To have a clear understanding of the correlation effect
in the CDW phase in ScV$_6$Sn$_6$, we performed a systematic density 
functional theory plus dynamical mean field theory (DFT + DMFT) 
calculations.
The resulting static local spin susceptibility is 
nearly independent of temperature, indicating the absence of local moment 
on atom V, in full agreement with experimental measurements.
The mass enhancements of quasiparticles and 
 bandwidth renormalizations near the Fermi level show a weak correlation 
strength in ScV$_6$Sn$_6$. In addition, the comparable mass enhancements of quasiparticles in ScV$_6$Sn$_6$ 
with CDW order and YV$_6$Sn$_6$ without CDW phase suggests that electronic 
correlations corresponding to Fermi surface nesting do not play the dominant role in 
the formation of CDW order in ScV$_6$Sn$_6$.
\end{abstract}

\pacs{Valid PACS appear here}
\maketitle
\section{INTRODUCTION}
Kagome lattice has received significant attention 
in last decades due to its fundamental
symmetry and physics, which often present as 
linear band crossing, saddle point, and flat 
band,  etc\cite{kagome1,kagome2,kagome3,kagome4,kagome5,kagome6,kagome7,kagome8}. 
In recent years, much effort has been devoted to the study of these 
characteristic features present by Kagome lattice in real materials 
and impressive progresses have been made in this direction. 
{  For example, linear band crossings and flat bands in
paramagnetic CoSn \cite{CoSn1,CoSn2} and antiferromagnetic 
FeSn \cite{FeSn1,FeSn2}, and saddle points 
generated van Hove singularity in antiferromagnetic YMn$_6$Sn$_6$ \cite{YMn6Sn6} 
and paramagnetic GdV$_6$Sn$_6$ \cite{GdV6Sn6}.}
{ These realistic materials provide practical platforms for the
study of the novel quantum phases, such as the anomalous Hall 
effect, topological band structure, flat bands, as well as
fractional quantum Hall effect, {  etc.}}

{  In addition, some other interesting properties were found
in Kagome metals. A series of Kagome metals in 
$A$V$_3$Sb$_5$ ($A$ = K, Rb, Cs) were discovered to exhibit 
superconductivity at low temperatures ranging from 0.9 to 
2.5 K \cite{135Tc1,135Tc2,135Tc3}. Moreover, 
charge density wave (CDW) transitions at temperatures between 80 and 
100 K were also observed in these 
systems\cite{tcdw1,135Tc1,135Tc2,135Tc3,tcdw5,tcdw6,tcdw7,tcdw8}, 
and van Hove singularity was believed to be one of the most important
prequirements \cite{vhs1,vhs2,vhs10,vhs11,vhs3,vhs4,vhs5,vhs6,vhs7,vhs8,vhs9,vhs12}}. 
Very recently, 
{  a long-range CDW order 
of $\textbf{\textit{q}}_{CDW}$ = (1/3, 1/3, 1/3) 
was observed in ScV$_6$Sn$_6$ below 92 K,
where the Kagome lattice constituted 
of atoms V \cite{166_CDW_EXP}, the same as $A$V$_3$Sb$_5$ ($A$ = K, Rb, Cs).}
However, the CDW order in ScV$_6$Sn$_6$ is mainly caused by the out-plane 
charge density modulation \cite{166_CDW_EXP}, which contrasts with 
$A$V$_3$Sb$_5$ (A = K, Rb, Cs) but coincide with FeGe. To date, 
several theoretical {  framework} to explore the origin 
of the CDW order in ScV$_6$Sn$_6$. According to current understanding, 
Fermi surface nesting (FSN) was ruled out by density functional theory 
(DFT) calculations \cite{Yanbinghai_PRL,wangnanlin} and a giant 
$q$-dependent electron-phonon coupling has been proposed as an 
underlying mechanism \cite{caochao_EPC}. However, it is important 
to consider another possible reason for electronic correlations 
arising from 3$d$ orbitals in the atoms V that were not taken 
into account. 
{  Therefore, to achieve a comprehensive 
understanding, it is necessary to investigate the effect 
of electronic correlations, particularly in the formation of 
CDW order in ScV$_6$Sn$_6$.}

In this work, we carried out a systematic study on ScV$_6$Sn$_6$ 
by using density functional theory plus dynamical mean field 
theory (DFT + DMFT) method. The calculated dependence of static 
spin susceptibility on temperature indicates the absence of 
local moment on atom V and we attribute it to the strong 
hybridization between V 3$d$ orbital and Sn 5$p$, Sc 3$d$ orbitals. 
The calculated mass enhancement $m^*/m_{DFT}$ is about 1.3 and 
insensitive to the variation of temperature, similar to 
{  that} in $A$V$_3$Sb$_5$ ($A$ = K, Rb, Cs) \cite{liumin,zhaojianzhou}. 
Meanwhile, there are some weak variations in band structure near 
the Fermi level ($E_F$) brought by electronic correlations. 
{  Hence, ScV$_6$Sn$_6$ is} recognized as a weakly 
correlated metal. Furthermore, the Fermi surface is slightly 
affected by electronic correlations, validating the conclusion 
that FSN is not the driving force of CDW order in ScV$_6$Sn$_6$ 
drawn by previous DFT calculations \cite{caochao_EPC,wangnanlin,Yanbinghai_PRL}. 
We also compared the mass enhancements in ScV$_6$Sn$_6$ and YV$_6$Sn$_6$ 
where no CDW order was observed. The comparable values suggest that 
electronic correlations are not crucial for determining the 
CDW order in ScV$_6$Sn$_6$. 

\section{METHODS}
We conducted fully charge self-consistent DFT + DMFT calculations 
{  by using} the eDMFT code \cite{haule2010dynamical} developed by 
Haule \textit{et al}. The DFT part employed the linearized 
augmented plane-wave method as implemented in the WIEN2K 
package \cite{blaha2001wien2k}. The Perdew-Burke-Ernzerhof 
generalized gradient approximation \cite{perdew1996generalized} 
was utilized for the exchange-correlation functional. 
Brillouin-zone integrations were performed on a 21 × 21 × 10 mesh. 
The atomic sphere radii ($R_{\rm MT}$) for Sc, V, and Sn were 
all set to 2.50 Bohr, and the plane-wave cutoff 
($K_{\rm max}$) was determined by $R_{\rm MT} \times K_{\rm max}=7.0$. 
In the DMFT calculations, the projector to the local Green's function was 
fixed to the solution of the Dirac equation at the DFT level. Hybridization 
functions were considered within an energy window of 20 eV centered around 
$E_F$. 
{  We deal with the double counting by the formula of $U(n-1/2)-J_H (n-1)/2$, 
where $n = 3.0$ represents the nominal occupation of V 3$d$ electrons,
$U$=5.0 eV represents the Coulomb interaction parameter, and 
$J_H$=0.7 eV represents the Hund's exchange parameter. The adopted values are consistent with previous 
calculations in V-based correlated compounds, including $A$V$_3$Sb$_5$ \
($A$ = K, Rb, Cs) \cite{liumin,zhaojianzhou}and vanadium oxides including 
SrVO$_3$ \cite{srvo3_1,srvo3_2}, CaVO$_3$ \cite{cavo3}, VO$_2$\cite{vo2} 
and V$_2$O$_3$\cite{v2o3_1,v2o3_2}.}
For the DMFT impurity solver, we 
adopted the density-density form of Coulomb repulsion, which significantly 
accelerated the calculations. All five V 3$d$ orbitals were treated as 
correlated. The DMFT quantum impurity problem was solved using the 
continuous-time quantum Monte Carlo (CTQMC) 
method \cite{werner2006continuous,haule2007quantum} at a temperature 
of $T$ = 116 K. Upon achieving the desired accuracy, analytical 
continuation was performed using the maximum entropy 
method \cite{haule2010dynamical} to obtain the self-energy on the real 
axis and subsequently calculate the electronic structure. We have 
confirmed that the incorporation of spin-orbit coupling has 
negligible influence on the conclusion. We therefore only present the 
results without considering spin-orbit coupling in the following. 

\section{RESULTS AND DISCUSSION}
\subsection{Crystal structure}
Figure 1(a) is the three-dimensional crystal structure of 
ScV$_6$Sn$_6$, {  showing a typical} HfFe$_6$Ge$_6$-type compound. 
Within one unit cell of ScV$_6$Sn$_6$, there are two Kagome layers 
formed by V and Sn2 atoms, one honeycomb layer formed by Sn3 atoms 
and one triangular layer formed by Sc and Sn1 atoms, see Fig. 1(b). 
Within each Kagome layer, Sn2 atoms are slightly off the plane formed 
by V atoms. 
{  The lattice constants and internal coordinates used in our 
calculations are optimized by the Vienna \textit{Ab initio} Simulation
Package (VASP) \cite{vasp}, until the atomic forces become less than 
10$^{-3}$ eV/$\rm \AA$.} The obtained lattice constants and internal 
atomic coordinates are given in Table I, which are in good agreement with 
experimental reports \cite{166_CDW_EXP} and previous 
calculations \cite{Yanbinghai_PRL}.

\begin{figure}
\includegraphics[width=3.5 in]{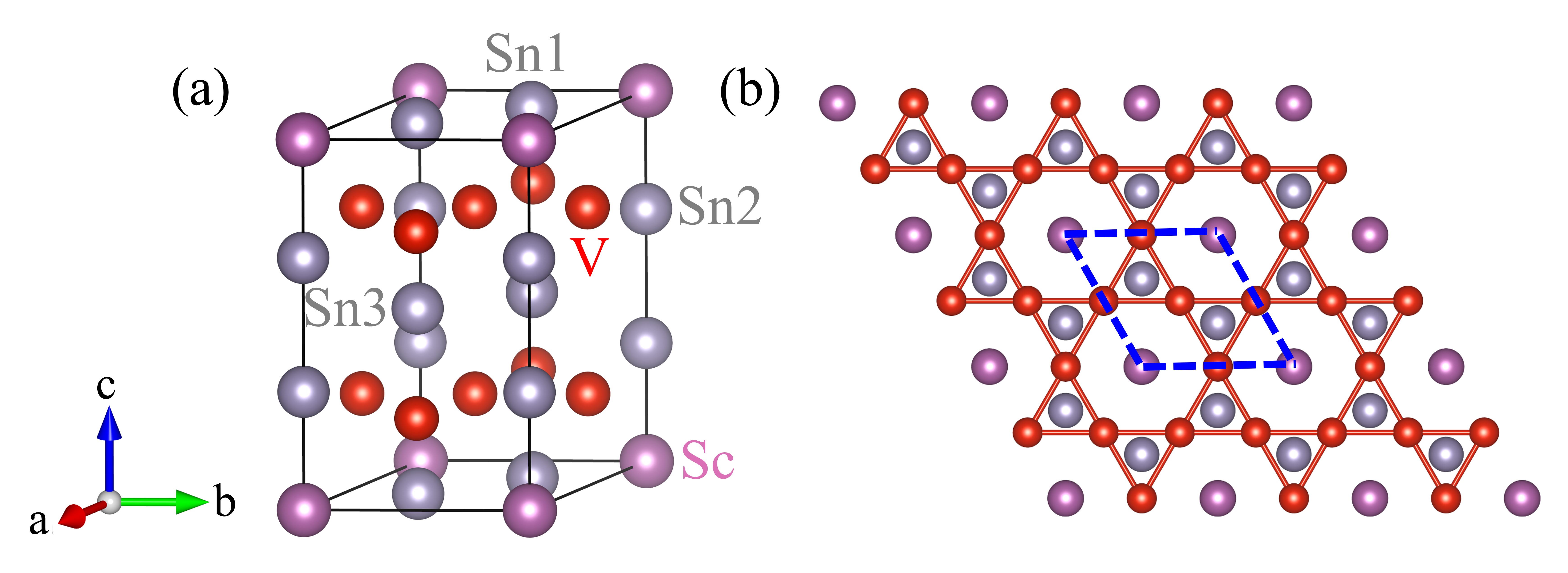}
\caption{\label{fig:wide}(a) Crystal structure of ScV$_6$Sn$_6$ 
with space group P6/mmm (No. 191). The gray, red, and pink atoms 
present atoms Sn, V, and Sc, respectively. (b) Top view along the c 
axis. Blue lines denote the unit cell of ScV$_6$Sn$_6$.}
\end{figure} 

\begin{table}[b]
\caption{\label{tab:table1}%
The relaxed internal atomic coordinates in ScV$_6$Sn$_6$ with the 
lattice constants a = b = 5.454 $\rm \AA$, and c = 9.230 $\rm \AA$.}
\begin{ruledtabular}
\begin{tabular}{lllll}
 Atom  & Wyckoff position & x & y & z \\
\hline
Sc & 1a  & 0  & 0 & 0\\
V & 6i  & 1/2 & 0 & 0.248 \\
Sn1 & 2c  &  1/3  & 2/3 &  0 \\
Sn2 & 2e & 0 & 0 & 0.320 \\
Sn3 & 2d & 1/3 & 2/3 & 1/2
\end{tabular}
\end{ruledtabular}
\end{table}

\subsection{Magnetism}

\begin{figure*}
\includegraphics[width=7.0in]{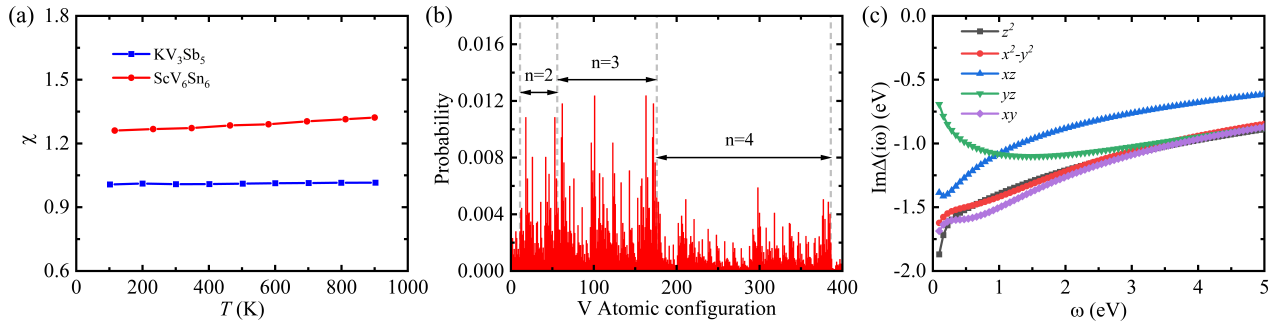}
\caption{\label{fig:wide}(a) 
{  Temperature dependent spin susceptibility 
obtained from CTQMC.} The blue and red dots are
ScV$_6$Sn$_6$ and KV$_3$Sb$_5$, respectively. The values of 
spin susceptibility of KV$_3$Sb$_5$ are from 
reference \cite{zhaojianzhou}. (b) Density functional theory 
(DFT) + dynamical mean field theory (DMFT) calculated atomic 
histogram of V 3$d$ orbitals in ScV$_6$Sn$_6$, with $U$ = 5 eV 
and $J_H$ = 0.7 eV. The states with occupation number exceeding 
five are not shown due to their negligible probabilities. 
(c) Imaginary part of the hybridization function on Matsubara 
frequencies at $T$ = 116 K.}
\end{figure*} 

To investigate the magnetism of ScV$_6$Sn$_6$ and 
compare our calculational results with available experimental 
results, we present the 
{  temperature dependent static local spin susceptibility}
in Fig. 2(a). The static local 
spin susceptibility { from} CTQMC approach is defined 
as $\chi=\int_{0}^{\beta}<S_z(\tau)S_z(0)>d\tau$, here $\beta$ 
is the inverse temperature and $\tau$ is the imaginary time. 
We take temperatures ranging from 58 to 900 K in the paramagnetic 
state in the calculations. In a magnetic system with local moments, 
this value would show a $T$ dependence, typically following the 
Curie-Weiss law. However, as presented by the red dots in Fig. 2(a), 
the susceptibility exhibits a paramagnetism with an almost flat 
line independence of $T$. For good metals, this behavior indicates the 
the dominance of a Pauli paramagnetic response from itinerant electrons 
and the absence of local moments. This paramagnetic behavior resembles 
another V-based Kagome metal KV$_3$Sb$_5$ \cite{zhaojianzhou}, as shown 
by the blue line in Fig. 2(a). 
{  The DFT + DMFT results here
is consistent with recent experimental measurement \cite{166_CDW_EXP}.}

{To   have more insight, we present the probability distribution 
for the different atomic configurations of V 3$d$ shell in Fig. 2(b).}
Within the DFT + DMFT scheme, each V impurity has 1024 3$d$ states. The atomic histogram 
refers to the probability of finding a V impurity in each atomic state. We note 
that only those states occupied by N = 2, 3, and 4 electrons have considerable 
probabilities. Hence the states with larger orbital occupation number 
(N $\textgreater$ 5) are not presented in Fig. 2(b). For the states with the 
same occupation number N, we arrange the states in descending order in accordance 
with their $|S_z|$ values. Note that in the atomic limit, the configuration with 
occupancy $n_d = 3$ should typically favor the high-spin state according to the 
Hund’s rule. However, this is not the case in Fig. 2(b). Here, the close competition 
between different occupancy and spin states indicates strong charge and spin 
fluctuations in paramagnetic ScV$_6$Sn$_6$. This can be attributed to the strong 
hybridization between V 3$d$, Sn 5$p$, and Sc 3$d$ orbitals. Figure 2(c) presents 
the imaginary parts of hybridization functions of the V 3$d$ orbitals, 
which indicate a substantial delocalization of these orbitals 
(larger than 0.5 eV), especially for V-$d_{z^2}$ orbital.

\begin{figure}[b]
\includegraphics[width=3.5in]{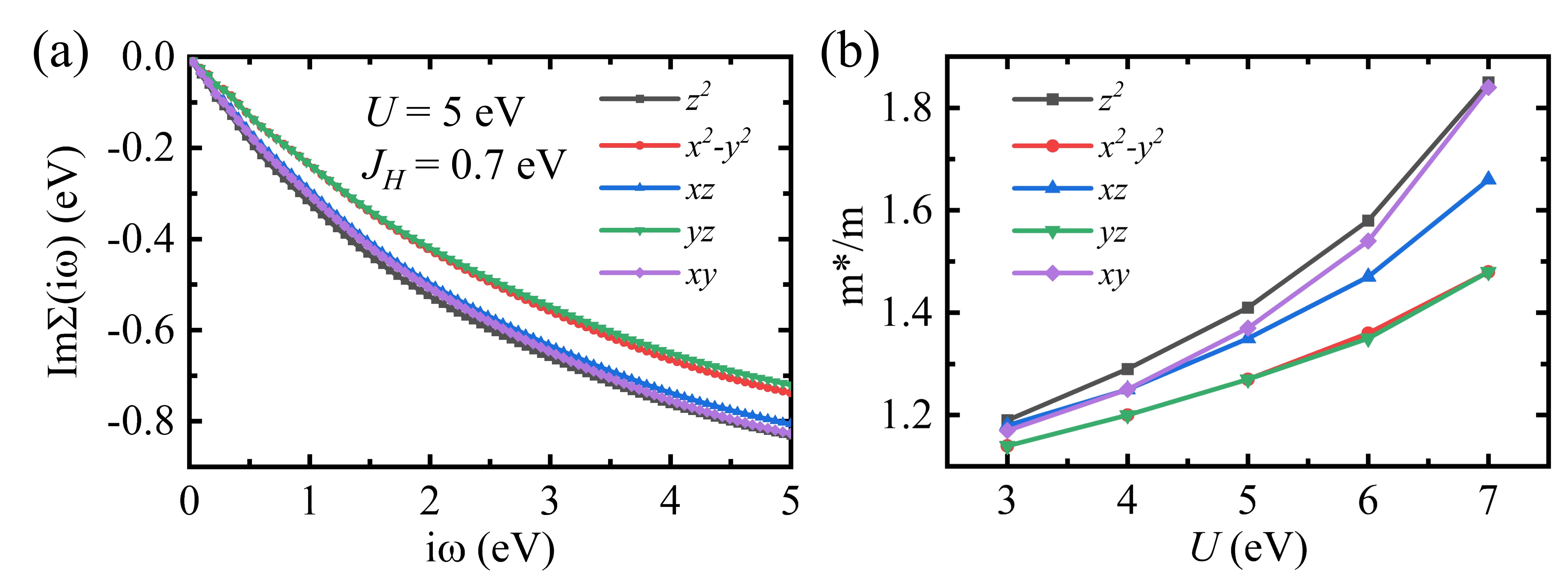}
\caption{\label{fig:wide}(a) Imaginary part of the self-energy of V 3$d$ 
shell on Matsubara frequencies at $T$ = 116 K. (b) The mass enhancement 
$m^*/m_{DFT}$ in ScV$_6$Sn$_6$ with different combinations of Coulomb 
interaction $U$ and $J_H$ (for each value of $U$, $J_H$ is equal to $0.14 \times U$).}
\end{figure} 

\subsection{Mass enhancement}
The effective mass enhancement is a straightforward and quantitative way 
to characterize the strength of electronic correlations. For weakly 
correlated systems, it is  almost equal to 1 but much greater than 1 
in strongly correlated systems, such as iron-based 
superconductors \cite{yinFe-based_SC} and heavy fermion 
materials \cite{heavy1,heavy2}. Within the DFT + DMFT method, 
we can obtain the effective mass enhancement $m^*/m_{DFT}$, 
which is equal to $1/Z$, and $Z$ is the quasiparticle weight
$Z=1/(1-\frac{\partial Im \Sigma(i \omega_n)}{\partial \omega_n}|_{\omega_n \rightarrow 0})$,   
according to the calculated self-energy on Matsubara frequencies.

Figure 3(a) presents the imaginary parts of self-energy for all five 
V 3$d$ orbitals at the temperature of 116 K in the paramagnetic state. 
{   
In practice, we obtain $m^*/m_{DFT}$ for each V 3$d$ orbital by 
average the values computed by origin and the lowest Matsubara 
point and the lowest two Matsubara points. }
Figure 3(b) displays 
the computed mass enhancement $m^*/m_{DFT}$ in ScV$_6$Sn$_6$ with 
different values of Coulomb interaction $U$ ranging from 3 to 7 eV 
and Hund interaction $J_H = 0.14\times U$. For a typical $U$ of 5.0 eV 
and $J_H$ of 0.7 eV that is commonly used for other V-based 
compounds \cite{srvo3_1,cavo3,liumin,srvo3_2,v2o3_1,v2o3_2,vo2,zhaojianzhou}, 
our computed values of $m^*/m_{DFT}$ in ScV$_6$Sn$_6$ at 116 K are 1.406, 
1.266, 1.345, 1.265, and 1.369 for V-$d_{z^2}$, V-$d_{x^2-y^2}$, V-$d_{xz}$, 
V-$d_{yz}$, and V-$d_{xy}$ orbitals, respectively. These values are significantly 
smaller than the typical values in strongly correlated systems, 
{  suggesting the weak correlations in this compound.} 
As shown in Figure 3(b), the V-$d_{z^2}$ orbital emerges as the most responsive 
to changes in the interaction parameters. However, despite setting the value 
of $U$ to 7 eV, the mass enhancement of V-$d_{z^2}$ orbital only reaches a 
maximum value of 1.850. Furthermore, it has been confirmed that the computed 
value of $m^*/m_{DFT}$ shows a small variation with temperature. For instance, 
the calculated $m^*/m_{DFT}$ of V-$d_{z^2}$ is 1.421 at 58 K and 1.400 at 232 K.

\subsection{Correlated electronic structure}

\begin{table*}[!htbp]
\caption{\label{tab:table2}%
DFT + DMFT calculated orbital occupation of V 3$d$ orbitals 
and mass enhancement obtained from the quasiparticle self-energy.}
\begin{ruledtabular}
\begin{tabular}{cccccccccccc}
  
  & \multicolumn{5}{c}{Orbital occupation} & &  \multicolumn{5}{c}{Mass enhancement} \\
  \cline{2-6}
  \cline{8-12}
   & $d_{z^2}$	&$d_{x^2-y^2}$	&$d_{xz}$	&$d_{yz}$	&$d_{xy}$ & &$d_{z^2}$	&$d_{x^2-y^2}$&	$d_{xz}$	&$d_{yz}$&	$d_{xy}$ \\
  \hline
  ScV$_6$Sn$_6$ & 0.803&	0.512&	0.569&	0.454&	0.924& &	1.406&	1.266&	1.345&	1.265&	1.369 \\
  \hline
  YV$_6$Sn$_6$ & 0.790&	0.500&	0.590&	0.459&	0.906& &	1.423&	1.275&	1.368&	1.271&	1.393 \\
\end{tabular}
\end{ruledtabular}
\end{table*}

To investigate the effect of electronic correlations on 
electronic structure of ScV$_6$Sn$_6$, we firstly plot the 
{ energy dispersion} and density of states (DOS) calculated by DFT in Fig. 4(a). 
There is a large peak in DOS around 0.3 eV above  $E_F$ with a V 3$d$ 
character, which could be attributed to the flat band alone the 
$\Gamma$-$M$-$K$-$\Gamma$ path, a typical feature 
present by Kagome lattice. From the band structure, we observe that 
this flat band is mainly contributed by V-$d_{z^2}$ orbital. Another 
two characteristics of Kagome lattice can also be seen, including 
linear band crossings at $K$ and saddle points at $M$ just below $E_F$,
in agreement with previous studies \cite{caochao_EPC,Yanbinghai_PRL}.

As a comparison, we plot 
the momentum-resolved spectral functions and momentum-integrated spectral 
functions obtained by the DFT + DMFT method in Fig. 4(b). The most obvious 
feature of the momentum-resolved spectral functions is the incoherence away 
from $E_F$, indicating a somewhat short quasiparticle lifetime away from $E_F$. 
However, the spectra near $E_F$ exhibit good coherence, and the bandwidths are 
weakly renormalized when compared with the DFT band structure represented by the 
gold dashed lines. Overall, our calculations show that ScV$_6$Sn$_6$ could be 
a weakly correlated metal.   {It is necessary to emphasize that 
while local correlation is adequately considered in the DFT + DMFT calculation, the 
influence of long-range correlation on the electronic structure of ScV$_6$Sn$_6$ 
has been omitted. Therefore, it is imperative to investigate the impact of 
long-range correlation in future research. For instance, employing the 
DFT + HSE06 method would provide valuable insights into this aspect.}

{  Figure 5(a) shows the DFT Fermi surface in the first Brillouin zone.} 
{  In contrast to $A$V$_3$Sb$_5$ ($A$ = K, Rb, Cs) \cite{liumin,zhaojianzhou},
one prominent distinguishing feature is the three-dimensionality with respect to $k_z$.}
Comparing the Fermi surfaces {  from} DFT and DFT + DMFT, 
one can see that the most obvious variation brought by 
electronic correlations are the increment of the size of the pocket 
centered at the middle of $L-M$, see Fig. 5(b).
This variation could be due to the lift of the bands forming a 
crossing near $E_F$ alone $L-M$. However, apart from this discrepancy, 
the Fermi surface obtained by the DFT + DMFT method is quite similar to that 
from DFT.

\begin{figure}[t]
\includegraphics[width=3.5in]{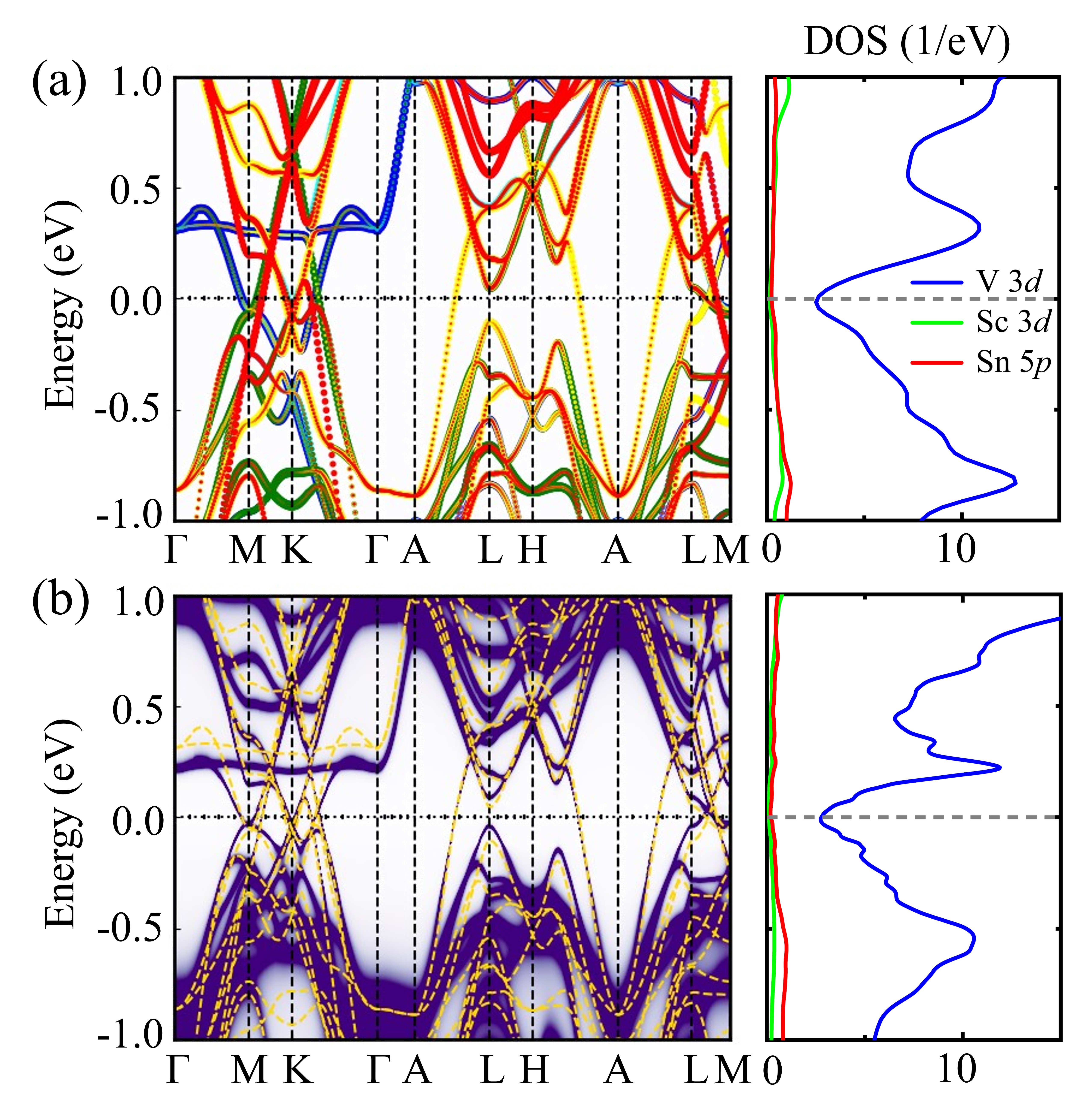}
\caption{\label{fig:wide} 
(a) {  Orbital projected DFT energy dispersion and density of states,
with V-$d_{z^2}$, V-$d_{x^2-y^2}$, V-$d_{xy}$, V-$d_{xz}$ and V-$d_{yz}$ orbitals 
labeled by blue, cyan, green, yellow and red lines, respectively.}
(b) Momentum-resolved spectral functions and momentum-integrated spectral 
functions from DFT + DMFT method at \textit{T} = 116 K. 
{  The DFT band structure given by gold dashed lines is
	attached for comparison.}}
\end{figure} 

\subsection{Charge density wave}

\begin{figure}[b]
\includegraphics[width=3.5 in]{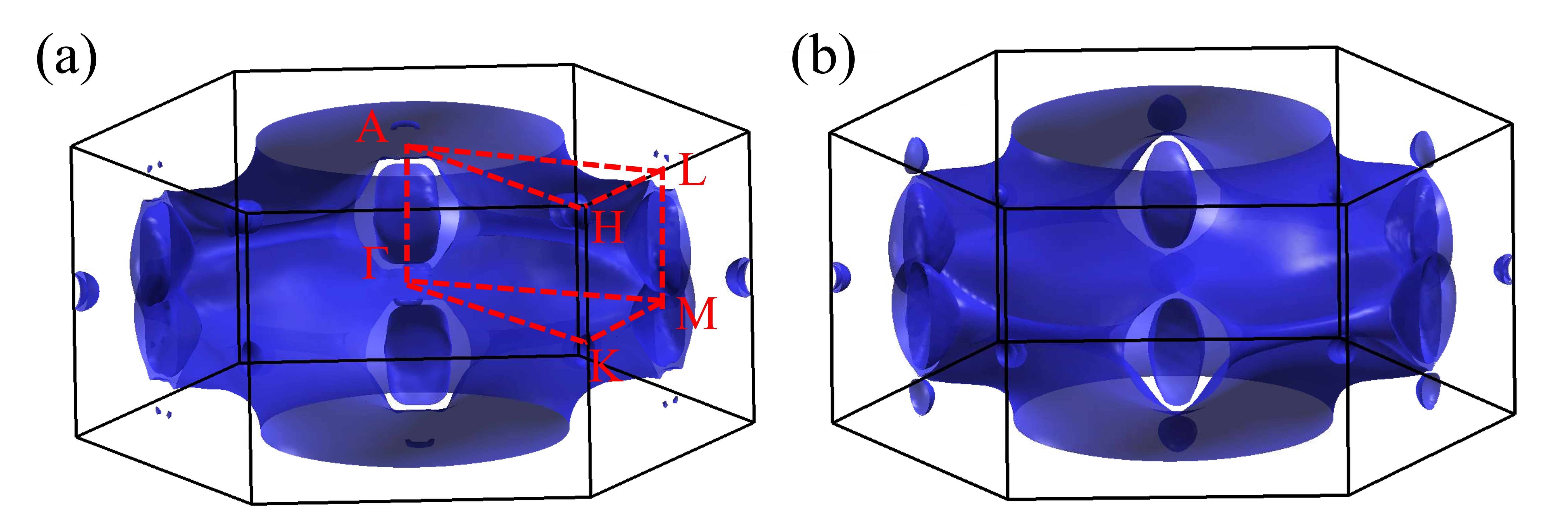}
\caption{\label{fig:wide}(a) DFT and (b) DFT + DMFT calculated Fermi surfaces 
of ScV$_6$Sn$_6$ in the first Brillouin zone.}
\end{figure} 

\begin{figure}[bht]
\includegraphics[width=3.5 in]{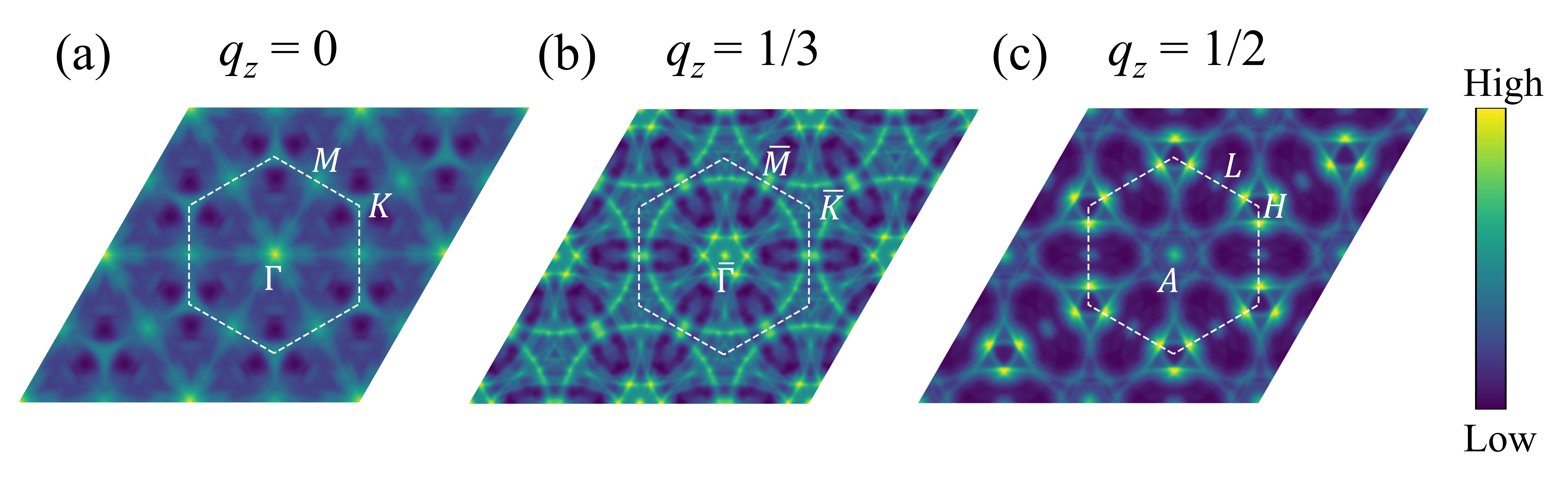}
\caption{\label{fig:wide}The DFT + DMFT calculated imaginary parts of the 
electron susceptibility $\chi(q)$, namely Fermi surface nesting function, 
on (a) $q_z = 0$, (b) $q_z = 1/3$ and (c) $q_z = 1/2$ planes. White hexagons 
in all panels represent the in-plane Brillouin zone, corresponding 
high-symmetry points on each plane are labeled.}
\end{figure} 

FSN is one of the underlying mechanisms of the lattice instability. 
The idea is that if Fermi surface contours coincide when shifted 
along the observed CDW wave vector, then the CDW is considered to 
be nesting derived. 
The imaginary part of the bare charge susceptibility 
under zero-frequency limit, namely FSN function, describing the degree of 
the FSN with a certain nesting vector $q$ is defined as
$\displaystyle \lim_{\omega \rightarrow 0}
\chi_0''(\bm{q},\omega)/\omega=\displaystyle \sum_{nn',\bm{k}}\delta(\epsilon_{n,\bm{k}}-\epsilon_0)\delta(\epsilon_{n',\bm{k+q}}-\epsilon_0)$
where $\delta$ is the Dirac-delta function, $\epsilon_{n,\bm{k}}$  is the eigenvalue 
of nth band at $k$ point, and $\epsilon_0$  is the Fermi energy. 

FSN was revealed as a strong candidate for driving CDW formation in 
rare-earth tritellurides, including GdTe$_3$ and LuTe$_3$ \cite{GdTe3}. 
However, according to previous DFT calculations, FSN was excluded as 
the primary driving force behind the CDW order in ScV$_6$Sn$_6$ \cite{Yanbinghai_PRL,wangnanlin}. 
To take electronic correlations into account, we plot the DFT + DMFT calculated FSN 
function on the $q_z$ = 0, 1/3, and 1/2 planes, respectively, in Fig. 6. Here a dense k-mesh 
of $100 \times 100 \times 51$ and a small Gaussian broadening factor of 0.001 eV 
in Dirac-delta functions are used to display the relative 
strength of the FSN function at different $q$ points. 
{  In contrast to $A$V$_3$Sb$_5$ ($A$ = K, Rb, Cs) with weak $q_z$ dependence \cite{liumin,zhaojianzhou}, the maximum of the FSN function of ScV$_6$Sn$_6$ undergoes a 
transition.} Specifically, as $q_z$ varies from 0 to 1/3 to 1/2, the maximum shifts 
from $M$ (Fig. 6(a)) to alone $\Gamma-M$ and near $\Gamma$ (Fig. 6(b)), and ultimately 
settles at the middle of $L-H$ (Fig. 6(c)). Single crystal x-ray diffraction determined CDW 
wave vector in ScV$_6$Sn$_6$ is $\textbf{\textit{q}}_{CDW}$ = (1/3, 1/3, 1/3)\cite{166_CDW_EXP}. 
If the CDW is indeed caused by FSN, the maximum should be located at the $\overline{K}$ point in 
Fig. 6(b), but this is not the case. We therefore conclude that FSN is 
unlikely to be the direct cause of the CDW instability in ScV$_6$Sn$_6$ 
even when electronic correlations are considered.

To further illustrate the dependence of CDW order on electronic correlations, 
we compare the correlation strength of quasiparticles in ScV$_6$Sn$_6$ and 
non-CDW compound YV$_6$Sn$_6$. Owing to the larger radius of 
atom Y, lattice constant a is larger in YV$_6$Sn$_6$ than ScV$_6$Sn$_6$ \cite{Y166}. 
However, as listed in Table II, the calculated orbital occupation numbers and 
mass enhancements in these two compounds exhibit a striking similarity. 
For instance, the mass enhancement of the most correlated orbital 
V-$d_{z^2}$ is 1.406 in ScV$_6$Sn$_6$ and 1.423 in YV$_6$Sn$_6$. 
We propose that electronic correlations are not crucial for the 
formation of CDW order in ScV$_6$Sn$_6$.

Recently, Cao \textit{et al}. \cite{caochao_EPC} observed a remarkably 
large $q$-dependent electron-phonon coupling, which has been proposed 
as the underlying mechanism responsible for the CDW order in ScV$_6$Sn$_6$. 
Given the weak electronic correlation strength observed in our DFT + DMFT 
calculations, the electronic correlations will not 
significantly affect this conclusion. Therefore, we support the notion that 
the $q$-dependent electron-phonon coupling is a strong candidate for driving 
the CDW order in ScV$_6$Sn$_6$.

\section{Conclusions}
In summary, based on DFT + DMFT {  calculations}, we studied the magnetism and 
electronic correlation effects in ScV$_6$Sn$_6$. 
{  The spin susceptibility is nearly independent of temperature, 
indicating the absence of local moment on atom V and consistent with experimental 
reports.} By analyzing the computed atomic histogram and hybridization 
functions, we attribute the absence of local moment to the strong 
hybridization between V 3$d$, Sn 5$p$, and Sc 3$d$ orbitals. 
The mass enhancement is about 1.3 and the bandwidths near $E_F$ 
are weakly renormalized by electronic correlations, indicating the 
weak electronic correlation strength in ScV$_6$Sn$_6$. The calculated 
FSN function with the inclusion of electronic correlations shows no 
maximum at experimentally determined $\textbf{\textit{q}}_{CDW}$, 
suggesting that FSN is not the driving force of CDW order in ScV$_6$Sn$_6$. 
{  Furthermore, the comparable values of quasiparticle mass 
enhancements in ScV$_6$Sn$_6$ and non-CDW YV$_6$Sn$_6$ suggest that 
the electronic correlations play a non-crucial role in the formation of 
the CDW order in ScV$_6$Sn$_6$.}

\begin{acknowledgments}

This work was supported by the National Key R\&D Program of China 
(Grant No. 2021YFB3501503) and the National Natural Science 
Foundation of China (Grants No. 52271016 and No. 52188101). 
Part of the numerical calculations in this study were carried out on 
the ORISE Supercomputer.

\end{acknowledgments}


%

\end{document}